\documentclass[12pt,letterpaper]{article}
{\setlength{\paperheight}{11in} \setlength{\paperwidth}{8.5in}}
\usepackage{graphicx}
\usepackage{graphics}
    
\begin{document}
\pagestyle{myheadings}
 \markright{Cryptography based on Composed Maps}
\title{Chaotic Cryptographic Scheme Based on Composition Maps}
\author{S. Behnia$^{a}$\thanks{E-mail:
S.behnia@iaurmia.ac.ir}, A. Akhshani$^{a}$, A. Akhavan$^{b}$, H. Mahmodi$^{a}$ \\
$^a${\small Department of Physics, IAU, Urmia, Iran.}\\
$^b${\small Department of Engineering, IAU, Urmia, Iran.}}
 \date{}
 \maketitle
 \begin{abstract}
In recent years, a growing number of cryptosystems based on chaos
have been proposed. But most of them encountered many problems such
as small key space and weak security. In the present paper, a new
kind of chaotic cryptosystem based on Composition of Trigonometric
Chaotic Maps is proposed. These maps which are defined as ratios of
polynomials of degree N, have interesting properties such as
invariant measure, ergodicity, variable chaotic region with respect
to the control parameters and ability to construct composition form
of maps. We have used a composition of chaotic map to shuffle the
position of image pixels. Another composition of chaotic map is used
in diffusion process. According to the performed analysis, the
introduced algorithm can satisfy the required performances such as
high level security, large key space and the acceptable encryption
speed.
\end{abstract}
\section{Introduction}
Chaos theory is established since 1970s from many different
research areas such as physics, mathematics, biology, engineering
and chemistry, etc. [Hao, 1993]. Chaotic systems have a number of
interesting properties such as ergodicity, the extreme sensitive
dependence on initial conditions, system parameters, mixing, etc.
Most properties are related to Shannon requirements of confusion
and diffusion  for constructing the cryptosystems[Shannon, 1949].
Due to tight relationship between chaos and cryptography [Brown
and Chua, 1996; kocarev et al., 1998; Alvarez et al., 1998;
Fridrich, 1998], there has been a great interest in developing
secure communication schemes utilizing chaos which protect
confidential information against eavesdropping and illegal access.
There exist two main approaches of designing chaos-based
cryptosystems: analog mode and digital mode. From 1989, along with
the use of analog chaotic systems in the design of secure
communication systems [Alvarez, 1999; Zhou and Ling, 1997; Lai et
al., 1999; Memon, 2003; Parlitz et al., 1992; Chen et al., 2003],
applications of computerized (also called digital) chaotic systems
in cryptography have attracted more and more attention [Baptista,
1998; Hong and Xieting, 1997; Jakimoski and Kocarev, 2001; Masuda
and Aihara, 2002; Matthews, 1989; Papadimitriou, 2001; Guan et
al., 2005; Xiao et al., 2005; Tang et al., 2005; Huang and Guan,
2005]. This paper chiefly focuses on the digital chaotic ciphers.
In the digital world nowadays, the security of digital images
becomes more important since the communications of digital
products over network occur more and more frequently. Thus, to
protect the content of digital images, some specific encryption
systems are needed. Due to some intrinsic features of images, such
as bulk data capacity and high correlation among pixels,
traditional encryption algorithms such as DES, IDEA and RSA are
not suitable for practical image encryption, especially under the
scenario of on-line communications. The main obstacle in designing
image encryption algorithms is that it is rather difficult to
swiftly permute and diffuse data by traditional means of
cryptology. In this respect, chaos-based algorithms have shown
their superior performance. By considering the advantages of
high-level efficiency and simplicity of one-dimensional chaotic
systems [Elnashaie and Abasha, 1995]. Different discrete-time
chaotic systems such as Logistic map used in image encryption
algorithms. Where there has been obvious drawbacks such as small
key space and weak security in introduced one-dimensional chaotic
cryptosystems [Kocarev, 2001; Ponomarenko and Prokhorov, 2002].\\
To eliminate these drawbacks. This paper aims to introduce a new
chaotic algorithm which has the advantages of high-level security,
large key space and the acceptable encryption speed. Since digital
images are usually represented as two-dimensional arrays, we
present algorithm based on Trigonometric Chaotic Maps and their
Composition [Jafarizadeh et al., 2001]. A diffusion process is
performed to confuse the relationship between cipher-image and
plain-image. By taking advantage of the exceptionally good
properties of Composition of Trigonometric Chaotic Maps ($CTCMs$)
such as mixing, sensitivity to initial conditions and system
parameters it was shown the proposed scheme incorporates $CTCMs$
and alternatively uses permutation and diffusion to transform the
image totally unrecognizable.\\ The remaining of the paper is
organized as follows. A brief description of $CTCMs$ is presented
in section 2. Section 3 presents the encryption algorithm based on
$CTCMs$. Some experimental results for verification are devoted in
section 4. In Section 5, security of the chaotic encryption
algorithm is perposed. Finally, Section 6 concludes the paper.
\section{Composition of Trigonometric Chaotic Maps}
We first review  the one parameter families of trigonometric chaotic
maps which are used to construct  the $CTCMs$. One-parameter
families of chaotic maps of the interval [0, 1] with an invariant
measure can be defined as the ratio of polynomials of degree N
[Jafarizadeh et al., 2001]:
\begin{equation}
\Phi_{N}^{(1,2)}(x,\alpha)=\frac{\alpha^2F}{1+(\alpha^2-1)F},
\end{equation}
Where {\bf F} substitute  with chebyshev polynomial of type one
$T_{N}(x)$ for $\Phi_{N}^{(1)}(x,\alpha)$ and chebyshev polynomial
of type two $U_{N}(x)$ for $\Phi_{N}^{(2)}(x,\alpha)$. We used
their conjugate or isomorphic maps. Conjugacy means that the
invertible map $ h(x)=\frac{1-x}{x}, $ maps $ I=[0,1]$ into $
[0,\infty) $ and transform maps $\Phi_{N}(x,\alpha)$ into
$\tilde{\Phi}_{N}(x,\alpha)$ defined as:
\begin{equation}
\tilde{\Phi}^{(1)}_{N}(x,\alpha)=\frac{1}{\alpha^{2}}\tan^{2}(N\arctan\sqrt{x}),
\label{eq:1}
\end{equation}
 \begin{equation}
\tilde{\Phi}^{(2)}_{N}(x,\alpha)
=\frac{1}{\alpha^{2}}\cot^{2}(N\arctan\frac{1}{\sqrt{x}}).
\label{eq:2}
  \end{equation}
 The map $\Phi_{2}^{(2)}(x,\alpha)$ is reduced to logistic one with $ \alpha=1$.
One can show that these maps have two interesting properties. The
first one is that $\Phi_{2}^{(1)}(\alpha,x)$ and
$\Phi_{4}^{(1)}(x,\alpha)$ maps have only one fixed point
attractor $x=1$ provided that their parameter depend on interval
$(2,\infty)$ and $(4,\infty)$. The second one is that at
$\alpha\geq 2$ and $\alpha\geq 4$ bifurcate to chaotic regime
without having any period
 doubling or period-n-tupling scenario and remain chaotic for all
$\alpha\in{(0,2)}$ and $\alpha\in{(0,4)}$  respectively. The map $
\Phi_{3}^{(1,2)}(x,\alpha)$ also has only one fixed point
attractor $x=0$ for $\alpha\in(\frac{1}{3}, 3)$. It bifurcates to
chaotic regime at $\alpha\geq\frac{1}{3}$, and remains chaotic for
$\alpha\in(0, \frac{1}{3})$.
 Finally it bifurcates at $\alpha=3$. When control parameter
 belong to $\alpha\in(\frac{1}{3}, \infty)$, then $x=1$ would be
 its corresponding fix point(see Fig. 1). From now on, depending on the situation,
  we will consider these maps Eqs. (2)-(3).\\
We have already derived analytically invariant measure for
One-parameter families of chaotic maps Eq. (1) by using arbitrary
values of the control parameter $\alpha$ and for each integer values
of $N$.
\begin{equation}
\mu_{\Phi_N^{(1,2)}(x,\alpha)}(x,\beta)=\frac{1}{\pi}\frac{\sqrt{\beta}}{\sqrt{x(1-x)}(\beta+(1-\beta)x)},\quad
\beta >0
\end{equation}
 With $\beta>0$ is the invariant measure of the
maps $\tilde{\Phi}_{N}^{(i)}(x,\alpha)$ provided that, we choose
the parameter $\alpha$ in the following forms:
\begin{equation}\alpha=\frac{\sum_{k=0}^{[\frac{N-1}{2}]}C_{2k+1}^{N}\beta^{-k}}{\sum_{k=0}^{[\frac{N}{2}]}C_{2k}^{N}\beta^{-k}}\end{equation}
in maps $\Phi_{N}^{(i)}(x,\alpha)$, N represents the odd values
and if N take even values, we would have the following equation:
\begin{equation}\alpha=\frac{\beta\sum_{k=0}^{[\frac{N}{2}]}C_{2k}^{N}\beta^{-k}}{\sum_{k=0}^{[\frac{N-1}{2}]}C_{2k+1}^{N}\beta^{-k}}\end{equation}
The symbol $[\quad]$ shows the greatest integer part [Jafarizadeh et
al., 2001 ].\\ Using the above hierarchy of family of one-parameter
chaotic maps we can generate a new hierarchy of families of
many-parameter chaotic maps with an invariant measure simply from
the composition of these maps. By the composition of maps Eqs.
(2)-(3), we can generate one-dimensional many-parameter chaotic
maps, which can be written in the following way:
\begin{equation}
\hspace{-0.2cm}\tilde{\Phi}_{N_1,.,N_n}^{\alpha_1,.,\alpha_n}(x)=
\frac{1}{\alpha_1^2}\tan^{2}\left(N_1\arctan\sqrt{
\frac{1}{\alpha_2^2}\tan^{2}(N_2\arctan\sqrt{..
\frac{1}{\alpha_n^2}\tan^{2}(N_n\arctan\sqrt{x})}..)}\right)
\label{eq:3}
\end{equation}
and
\begin{equation}
\hspace{1cm}\tilde{\Phi}_{N_1,.,N_n}^{\alpha_1,.,\alpha_n}(x)=
\frac{1}{\alpha_1^2}\cot^{2}\left(N_1 \arctan \frac{1}{\sqrt{
\frac{1}{\alpha_2^2}\cot^{2}(N_2 \arctan \frac{1}{\sqrt{..
\frac{1}{\alpha_n^2}\cot^{2}(N_n \arctan
\frac{1}{\sqrt{x}}}..)})}}\right)
\end{equation}

One can show that the chaotic regions are:
$\prod_{k=1}^{n}\frac{1}{N_k}< \prod_{k=1}^{n}\alpha_k
<\prod_{k=1}^{n} N_k $ for odd integer values of
$N_1,N_2,\ldots,N_n$. If one of the integers happens to become
even, then the chaotic region in the parameter space can be
defined by $\alpha_k>0,\mbox{for\quad} k=1,2,\ldots,n$ and
 $\prod_{k=1}^{n}\alpha_k <\prod_{k=1}^{n} N_k $. Out of these
 regions, they have only period one stable fixed points.
 The introduced maps Eqs. (7)-(8) follows the same measure Eq. (4). The
relation between the control parameters of composed maps and
$\beta$ is presented in our previous paper [Jafarizadeh and
Behnia, 2002].
\subsection{Ergodicity} It has been noticed that there exists an
interesting relationship between chaos and cryptography: many
properties of chaotic systems have their corresponding
counterparts in traditional cryptosystems, such as: Ergodicity and
Confusion, Sensitivity to initial conditions/Control parameter and
Diffusion. In cryptographic terms, ergodicity claims that it is
very hard to predict the actual position of a point from its
initial position. Moreover, after experiencing enough iterations,
every position within the whole block is equally likely to be the
actual position for almost every starting point (confusion).\\ A
transformation T is ergodic, if it has the probability that for
almost every $\omega$, the orbit $\{\omega, T\omega, T^{2}\omega,
...\}$ of $\omega$ is a sort of replica of $\Omega$ itself.
Formally, we shall say that T is ergodic if each invariant set A,
i.e.; a set such that T$^{-1}$(A)=A, is trivial in the sense that
it has measure either zero or one. T$^{-1}$(A)=A
$\Rightarrow$$\mu$(A)=0 or $\mu$(A)=1.
 In a
non-ergodic system for counter image set of A$\subset$ [0,1] we
have:
$$T^{-1}(A)=\{x\in[0,1] | y=T(x), y\in A\}$$
and the map is non-ergodic if  $0<\mu(A)<1$,  i.e., the invariant
measure which is not equal to zero or one, appears to be
characteristic of non-ergodic behavior [Medio, 1999]. Therefore,
the study, based on invariant measure analysis, can be useful for
confirming the ergodicity behavior of a map.
\subsection{Lyapunov characteristic exponent}
The Lyapunov exponent $\lambda$ provides the simplest information
about chaoticty. It can be computed by considering the separation of
two nearby trajectories evolving in the same realization of the
random process as follow [Dorfman, 1999]: $$
\lambda(x_{0})=lim_{n\rightarrow\infty}\frac{1}{n}\sum_{k=0}^{n-1}\ln\mid
\frac{d\tilde{\Phi}(x ,\alpha)}{dx}\mid,
$$
where $x_{k}=\overbrace{\tilde{\Phi}_{N} \circ \tilde{\Phi}_{N}\circ
....\circ \tilde{\Phi}_{N}^{k}(x_{0})}$. It is obvious that its
negative values, show that the system is under influence of fix
point (attractor) and its positive values show that the  system
follows repeller [Dorfman, 1999]. Also, the lyapunov number is
independet of initial point, provided that the motion inside the
invariant manifold is ergodic. Thus $\lambda{(x_{0})}$ characterizes
the invariant manifold of chaotic maps as a whole. In chaotic
region, chaotic maps are ergodic as Birkhof ergodic theorem predicts
[keller, 1998]. In non-chaotic region of the parameter, lyapunov
characteristic exponent is negative definite, since in this region,
we have only single period fixed points without bifurcation. Now for
composition of chaotic maps Eqs. (7)-(8):
\begin{equation}
\lambda_{N_1,\cdots,N_n}^{\alpha_1,\cdots,\alpha_n}(x_{0})
=\lim_{n\rightarrow\infty}\frac{1}{n}\sum_{k=0}^{n-1}\ln\mid\frac{d\tilde{\Phi}_{N_1,\cdots,N_n}^
{\alpha_1,\cdots,\alpha_n}(x_{k},\alpha)}{dx}\mid, \label{eq:4}
\end{equation}
where
$x_{k}=\overbrace{\tilde{\Phi}_{N_1,\cdots,N_n}^{\alpha_1,\cdots,\alpha_n}
\circ....\circ\tilde{\Phi}_{N_1,\cdots,N_n}^{\alpha_1,\cdots,\alpha_n}}
$. Thus
$\lambda_{N_1,\cdots,N_n}^{\alpha_1,\cdots,\alpha_n}(x_{0})$
characterizes the invariant manifold of
$\tilde{\Phi}_{N_1,\cdots,N_n}^{\alpha_1,,\cdots,\alpha_n}$ as a
whole. Therefore, these maps are ergodic in certain region of
their parameters space as explained above. In the complementary
region of the parameters space they have only a single period one
attractive fixed point. Also in contrary to the most of usual
one-dimensional one-parameter or many-parameters family of maps
they have only a bifurcation from a period one attractive fixed
point to chaotic state or vice-versa.
\section{Encryption Algorithm }
A possible way to describe the key space might be in terms of
positive Lyapunov exponents. By considering the Lyaponuv exponent of
one-dimensional many-parameter chaotic maps Eq. (9), we choose a
chaotic region of $CTCMs$ (see Figs. 2 (a)-(b)). In order to show
the capability of introduced model, we have used two simple models
of one dimensional two-parameter chaotic maps for generating
cryptosystem. By choosing (N$_{1}$=3, N$_{2}$=5) in Eq. (7):
\begin{equation}
 \Phi_{N_{1},N_{2}}^{\alpha_1,
\alpha_2}=\frac{1}{{\alpha_{{2}}}^{2}}\tan^{2} \left( N_{2}\,\arctan
\left( \sqrt {{\frac {  \tan^{2}
 \left( N_{1}\,\arctan \left( \sqrt {x} \right)  \right) }{{
\alpha_{{1}}}^{2}}}} \right)  \right) \hspace{0.29cm} \mbox{CTCM I}
\label{eq:6}
\end{equation}
and by considering (N$_{1}$=4, N$_{2}$=8) in Eq. (8), we have:
\begin{equation}
\Phi_{N_{1},N_{2}}^{\alpha^{'}_1, \alpha^{'}_2}=
\frac{1}{{\alpha^{'}_{{2}}}^{2}} \cot^{2} \left( N_{2}\,\arctan
\left( {\frac {{{\alpha^{'}_{{1}}}}}{\sqrt { {
  \cot^{2} \left( N_{1}\,\arctan \left( {\frac {1}{\sqrt {x}}} \right)
  \right)}}}} \right)  \right)
\hspace{0.5cm} \mbox{CTCM II} \label{eq:7}
\end{equation}
In order to encrypt the image, we have to go through both
permutation and XOR-ing processes. To permute the image, the
points are rearranged in the following way:
\[
\begin{array}{lc}
\mbox{Permutation}: \left\{ \begin{array}{rllll}
m: &  x_{0}, & \alpha_{1x}, & \alpha_{2x}, \\
n: & y_{0}, & \alpha_{1y}, & \alpha_{2y},
\end{array} \right.,\hspace{0.2cm}
 \end{array}
\]
XOR-ing is done in two stages. The first stage includes:
\[
 \mbox{XOR-ing Stage I}\left\{ \begin{array}{lcll}
 m\times n: x_{0},& \alpha_{1},&\alpha_{2},&
\end{array} \right.
\]
and the next step:
\begin{center}
$\mbox{XOR-ing Stage II}: \left\{ \begin{array}{rllll} m \times n: &
x_{0}, & \alpha^{'}_{1}, & \alpha^{'}_{2}
\end{array} \right.
$
\end{center}
 By choosing image M$_{m \times n}$
with $m \times n$ pixels, the encryption process can be explained
with the block diagram (Fig. 3). The image encryption can be done
through the following steps:
\begin{itemize}
  \item \emph{Step 1}: According to the following relations, with the help of $CTCM
  $ I, image $M_{m \times n}$ can be permuted by swapping the pixels:
\begin{equation}
x_{p}= \lfloor \phi_{N_{1}} \times 10^{14} \rfloor \hspace{0.2cm}
mod \hspace{0.2cm} 256
\end{equation}
\begin{equation}
y_{p}= \lfloor \phi_{N_{2}} \times 10^{14} \rfloor \hspace{0.2cm}
mod \hspace{0.2cm} 256
\end{equation}
\item  \emph{Step 2}: To diffuse the image
with using XOR possess, we use $CTCM$ II in the permuted image.
The generated result is stored in C$_{m \times n}$:
\begin{equation}
X_{k}=\lfloor x \times 10^{14} \rfloor \hspace{0.5cm} mod
\hspace{0.5cm} 256 \label{eq:8}
\end{equation}
\begin{equation}
C_{ij}=X_{k} \hspace{0.2cm} XOR \hspace{0.2cm}
\{{(M_{ij}+X_{k})\hspace{0.2cm} mode \hspace{0.2cm} 256}\}
\hspace{0.2cm} XOR \hspace{0.2cm} C_{p}
\end{equation}
 where C$_{p}$ is the modified previous pixel.
 \item\emph{Step 3}: This step requires another XOR possess in the
results of step 2 by using $CTCM$ I , Eq. (15) and the output
 encrypted image is known as a ciphertext.
\begin{equation}
E_{ij}=[\hspace{0.1cm}(x_{ij} \times 10^{14}) \hspace{0.2cm} mod
\hspace{0.2cm} 256] \hspace{0.2cm}  XOR \hspace{0.25cm} C_{ij}
\label{eq:9}
\end{equation}
\end{itemize}
For decryption the encrypted image one needs to receive encryption
keys and follow the introduced steps in reverse order. In decryption
process, we use the inverse of Eq. (15) which is introduced as
follows:
\begin{equation}
M_{ij}=\{ X_{k} \hspace{0.2cm} XOR \hspace{0.2cm} C_{p}
\hspace{0.2cm} XOR \hspace{0.2cm} C_{ij} + 256 -X_{k}\}
\hspace{0.2cm} mod \hspace{0.2cm} 256
\end{equation}
\section{Experimental Results} First take the encryption key,
then we implement the introduced model of encryption  on sample
image ('Boat' of size 256$\times$256) as our plain-image Fig.
4(a). The encrypted image is presented in Fig. 4(b). Where we have
used Visual C++ running program in a personal computer with 2.4
GHz Pentium IV, 256 Mb memory and 80 Gb hard-disk capacity. The
average time used for encryption/decryption on 256 grey-scale
images of size 256$\times$256 is shorter than 0.4s. In order to
encrypt the image, permutation and XOR-ing process are done in the
following way:
\[
\hspace{-0.5cm} \mbox{Permutation:} \left\{ \begin{array}{lcl}
x:\left\{\begin{array}{rlll}
\alpha_1=2.10155, & \alpha_2=3.569221, &  x_{0}=25.687, \\
\end{array}
\right. \\ \\
y: \left\{\begin{array}{rlll} \alpha_1=1.8874, & \alpha_2=4.23562, &
y_{0}=574.461,\\ \end{array} \right.
\end{array} \right.
\]
\begin{center}
$
 \mbox{XOR-ing Stage I:}\left\{\begin{array}{lcll}
 m\times n:\alpha_{1}=2.8912, & \alpha{2}=3.89954, &  x_{0}=814.217217,
\end{array} \right.
$
\end{center}
\[
\hspace{-0.5cm} \mbox{XOR-ing Stage II:} \left\{ \begin{array}{lcll}
 m\times n: \alpha^{'}_{1}=61.522, & \alpha^{'}_{2}=257.26223, &  x_{0}=79.82,
\end{array} \right.
\]
As shown by Figs. 4(c)-(d), the decryption process with wrong keys
\linebreak ($x_{0}$=2.10155400000001 in permutation and
$\alpha^{'}_{1}$=61.52200000000005 in XOR-ing) generates an image
with a random behavior. The sensitivity to initial conditions, which
is the main characterization of chaos, guarantees the security of
our scheme. The preformed experiments results show that the new
algorithm validly solves problem of encryption failure
caused by the small key space and weak security.\\
Statistical analysis, performed on the proposed image encryption
algorithm, demonstrates superior confusion and diffusion properties
of the algorithm. So, strongly resists statistical attacks. One
typical example is shown in Figs. 5(a)-(b).  Fig. 5(b), shows the
histogram of the ciphered image where it is fairly uniform and
significantly different with respect to the histogram of the
original image.
\section{Security Analysis}
A good encryption scheme should resist all kinds of known attacks,
such as\\ known-plain-text attack, cipher-text only attack,
statistical attack, differential attack, and various brute-force
attacks. Some security analysis have been performed on the
proposed image encryption scheme, including the most important
ones like key space analysis, information entropy and statistical
analysis which have demonstrated the satisfactory security of the
new scheme.
\subsection{Key space analysis} Key space size is the total number of
different keys that can be used in the encryption system. As
mentioned above, the key of the cryptosystem in the introduced
algoritem is composed of three parts: permutation parameters,
XOR-ing stage I and XOR-ing stage II parameters. Key space size in
our introduced example Eqs. (10)-(11) consists of 8 control
parameter and 4 initial conditions. As it was shows in Figs.
4(c)-(d) cryptosystem is completely sensitive to secret keys. If the
precision will be $10^{-14}$, the key space size for just initial
conditions is $10^{14\times4}=10^{56}\approx2^{186}$. It is
nessesery to remmember that the general model Eqs. (7)-(8) allows us
to increase the key space size with respect to level of security.
Therefore, the key space is very large and can resist all kinds of
brute-force attacks.
\subsection{Information entropy}
\label{}Information theory is a mathematical theory of data
communication and storage founded in 1949 by Claude E. Shannon. To
calculate the entropy \textit{H(s)} of a source \textit{s}, we have:
\begin{equation}
H(s)=\sum_{i=0}^{2N-1}P(s_{i})\log_{2} \frac{1}{P(s_{i})},
\label{eq:10}
\end{equation}
where $P(s_{i})$ represents the probability of symbol $s_{i}$.
Actually, given that a real information source seldom transmits
random messages, in general, the entropy value of the source is
smaller than the ideal one. However, when these messages are
encrypted, their entropy should ideally be 8. If the output of such
a cipher emits symbols with an entropy of less than 8, then there
exists a predictability which threatens its security. We have
calculated the infomation entropy for encrypted image Fig. 4(b):
\[
H(s)=\sum_{i=0}^{255}P(s_{i})\log_{2}\frac{1}{P(s_{i})}=7.997
\]
The obtained value is very close to the theoretical value 8.
Apparently, comparing it with the other existing algorithms, such
as [Xiang, 2006], the proposed algorithm is much more closer to
the ideal situation. This means that information leakage in the
encryption process is negligible, and so the encryption system is
secure upon the entropy attack.
\subsection{Correlation of two adjacent pixels}
To test the correlation between two adjacent pixels in plain-image
and ciphered image, the following procedure was carried out. We
randomly selected 1000 pairs of two adjacent (in vertical,
horizontal, and diagonal direction) pixels from plain-image and
ciphered image. Then we calculated the correlation coefficients
[Chen, 2004], respectively (see Table~\ref{tab:1} and Figs. 6(a)-(b)
by using the following two formulas:
\begin{equation}
cov(x,y)=\frac{1}{N}\sum_{i=1}^{N} (x_{i}-E(x))(y_{i}-E(y)),\quad
r_{xy}=\frac{cov(x,y)}{\sqrt{D(x)}\sqrt{D(y)}}, \label{eq:14}
\end{equation}
where
$$
E(x)=\frac{1}{N}\sum_{i=1}^{N} x_{i}, \quad
D(x)=\frac{1}{N}\sum_{i=1}^{N} (x_{i}-E(x))^2. \label{eq:12}
$$
$E(x)$ is the estimation of mathematical expectations of x, $D(x)$
is the estimation of variance of x and $cov(x,y)$ is the
estimation of covariance between x and y. where x and y are
grey-scale values of two adjacent pixels in the image.
\subsection{Differential attack}
 To test the influence of one-pixel change on the whole
 encrypted image by the proposed algorithm, two common measures
were used: NPCR and UACI [Chen and Ueta, 1999]. The number of
pixels change rate (NPCR) have been measured to see the influence
of changing a single pixel in the original image on the encrypted
image. The unified average changing intensity (UACI) measures the
average intensity of differences between the plain-image and
ciphered image. We take two encrypted images, C$_{1}$ and C$_{2}$,
whose corresponding original images have only one-pixel
difference. We label the grey scale values of the pixels at grid
(i,j) of $C_{1}$ and C$_{2}$ by C$_{1}$(i,j) and C$_{2}$(i,j), and
$C_{1}$ and $C_{2}$ have the same size. Then, D(i,j) is determined
by $C_{1}$(i,j) and C$_{2}$(i,j), that is, if $C_{1}$(i,j)=
$C_{2}$(i,j), then , D(i,j)=1 ; otherwise, D(i,j)=O. NPCR and UCAl
are defined by the following formulas:
\begin{equation}
NPCR=\frac{\sum_{i,j}D(i,j)}{W\times H} \times 100\%
\end{equation}
\begin{equation}
UACI=\frac{1}{W \times H}\left[\sum_{i,j}\frac{|C_{1}(i,j) -
C_{2}(i,j)|}{255}\right] \times 100\%
\end{equation}
Where, W and H are the width and length of the image. We obtained
NPCR=0.41751\% and UCAI=0.3314\%. With regard to obtained results,
it seems that the proposed algorithm has a good ability to resist
differential attack.
\section{Summery and Conclusion}
\label{6} In this paper, we propose a new scheme based on the
hierarchy of one dimensional chaotic maps of interval[0,$\infty$).
The chaotic properties such as mixing and sensitive dependence on
initial conditions and control parameters are suitably utilized
while the limitation and weaknesses of the chaotic encryption
system are effectively overcome. We have used composition form of
chaotic maps in order to increase both the number of keys (control
parameters) and complexities involved in the algorithm. By using
the composition of chaotic maps, we can increases the confusion in
the encryption process. It should be mentioned that increasing the
confusion in encryption results in increasing security in
cryptosystem. Furthermore, the introduced cryptosystem is very
robust to attacks whether it is based on statistical or reasoning
analysis. As it was shown by differential attack on encrypted
image, the system is also very sensitive with respect to the small
changes in the plaintext. According to the performed analysis, the
algorithm can satisfy most of the performances required such as
high level of security, large key space and the acceptable
encryption speed. Our presented cryptosystem is of practicality
and reliable value having to be adopted for Internet image
encryption, transmission applications, secure commination and
other information security fields.
\newpage
\section*{References}
\noindent Alvarez, G. Monotoya, F. Romera, M. Pastor, G. [1998]
"Chaotic cryptosystems," {\em In Proc. IEEE Int. Carnahan
Conf. Security Technology}, pp.332-338.\\
\noindent Baptista, M. S. [1998] "Cryptography with chaos," {\em Phys. Lett. A} \textbf{240}, 50-54.\\
\noindent Brown, R. and Chua, L. O. [1996] "Clarifying chaos:
examples and counterexamples," {\em Int. J. Bifurcation and
Chaos} \textbf{6}, 219-242.\\
\noindent Chen, G. Mao, Y. Chui, C. K. [2004] "A symmetric image
encryption scheme based on 3D chaotic cat maps," {\em Chaos
Solitons and Fractals} \textbf{21}, 749-761.\\
\noindent Chen, J. Y. Wong, K. W. Cheng, L. M.  Shuai, J. W.
[2003] "A secure communication scheme based on the phase synchronization of chaotic systems," {\em Chaos} \textbf{13}, 508-514.\\
\noindent Chen, G. and Ueta, T. [1999] "Yet another
chaotic attractor," {\em Int. J. Bifurcation and Chaos} \textbf{9}, 1465-1466.\\
\noindent Dorfman, J. R. [1999] "An Introduction to Chaos in
Nonequilibrium Statistical Mechanics," {\em Cambridge, Cambridge university press.}\\
\noindent Elnashaie, S. S. E. H. and Abasha, M. E. [1995] "On the
chaotic behaviour of forced fluidized bed catalytic reactors,"
{\em Chaos Solitons and Fractals} \textbf{5}, 797-831.\\
\noindent Fridrich, J. [1998] "Symmetric ciphers based on
two-dimensional chaotic maps," {\em Int. J. Bifurcation and
Chaos} \textbf{8(6)}, 1259-1284.\\
\noindent Guan, Z.-H. Huang, F. Guan, W. [2005] "Chaos-based
image encryption algorithm," {\em Phys. Lett. A} \textbf{346}, 153-157.\\
\noindent Hao, B.-L. [1993] "Starting with Parabolas: An
Introduction to Chaotic Dynamics" {\em Shanghai Scientific and
Technological Education Publishing House," Shanghai, China.}\\
\noindent Hong, z. and Xieting, L. [1997] "Generating chaotic
secure sequences with desired statistical properties and high
security," {\em Int. J. Bifurcation and Chaos} \textbf{7(1)}, 205-213.\\
\noindent Huang, F. and Guan, Z.-H [2005] "Cryptosystem using
chaotic keys," {\em Chaos Solitons and
Fractals} \textbf{23}, 851-855.\\
\noindent Jakimoski, G. and  Kocarev, L. [2001a] "Chaos and
cryptography: Block encryption ciphers based on chaotic maps,"
{\em
IEEE Trans.Circuits and Systems-I}, \textbf{48(2)}, 163-169.\\
\noindent Jafarizadeh, M. A. Behnia, S. Khorram, S. Nagshara, H.
 [2001] "Hierarchy of chaotic maps with an invariant measure," {\em J, Stat. Phys.} \textbf{104(516)}, 1013-1028.\\
\noindent Jafarizadeh, M. A. Behnia, S. [2002] "Hierarchy of
chaotic maps with an invariant measure and their compositions "
{\em J. Nonlinear Math.Phys.} \textbf{9(1)}, (2002), 1-16.\\
\noindent kocarev, L. Jakimoski, G.  Parlitz, U. [1998] "From
chaotic maps to encryption schemes," {\em in Proc.IEEE
Int.Symposium
Circuits and Systems} \textbf{4}, 514-517.\\
\noindent Kocarev, L. [2001] "Chaos-based cryptography: a brief
overview," {\em IEEE Circuits and Systems Magazine} \textbf{1(3)},
6-21.\\
\noindent keller, G. [1998] " Equilibrium State in Ergodic
Theory," {\em Cambridge, Cambridge university press.}\\
\noindent  Lai, Y.-C. Bollt, E. Grebogi, C. [1999] "Communicating
with chaos using two-dimensional symbolic dynamics," {\em Phys.
Lett. A} \textbf{255}, 75-81.\\
\noindent Masuda, N. and Aihara, K. [2002] "Cryptosystems with
discretized chaotic maps," {\em IEEE
Trans.Circuits and Systems-I} \textbf{49(1)}, 28-40.\\
\noindent Matthews, R. [1989] "On the derivation of a `chaotic'
encryption algorithm," {\em  Cryptologia}
\textbf{XIII(1)}, 29-42.\\
\noindent Medio, A. [1999] "Chaotic Dynamics: Theory and
Applicatios to Economics," {\em Cambridge, Cambridge University Press.}\\
\noindent Memon, Q. [2003] "Synchronized chaos for network
security," {\em Computer. Communications} \textbf{26},
498-505.\\
\noindent Parlitz, U.  Chua, L. O. Kocarev, L.  Halle, K. S.
Shang, A. [1992] "Transmission of digital signals by chaotic
synchronization," {\em Int. J. Bifurcation and Chaos} \textbf{2},
973-977.\\
\noindent Papadimitriou, S. Bountis, T. Mavroudi, S. Bezerianos,
A. [2001] "A probabilistic symmetric encryption scheme for very
fast secure communication based on chaotic systems of difference
equations," {\em Int. J. Bifurcation and Chaos} \textbf{11(12)},
3107- 3115.\\
\noindent Ponomarenko, V. I. and Prokhorov, M. I. [2002]
"Extracting information masked by the chaotic signal of a
time-delay system,"
{\em Phys. Rev. E} \textbf{66}, 026215\\
\noindent Shannon, C. E. [1949] "Communication theory of
secrecy systems," {\em Bell System Technical J.} \textbf{28}, 656-715.\\
\noindent Tang, G. Liao, X. Chen, Y. [2005] "A novel method for
designing S-boxes based on chaotic maps," {\em Chaos
Solitons and Fractals} \textbf{23}, 413-419.\\
\noindent Xiang, T. Liao, X. Tang, G. Chen, Y. Wong, K.-W.
[2006] "A novel block cryptosystem based on iterating a chaotic map," {\em Phys. Lett. A} \textbf{349}, 109-115.\\
\noindent Xiao, D. Liao, X. Wong, K. [2005] "An efficient entire
chaos-based scheme for deniable authentication," {\em Chaos
Solitons and Fractals} \textbf{23(4)}, 1327-1331.\\
\noindent Zhou, H. and Ling, X.-T. [1997] "Problems with the
chaotic inverse system encryption approach," {\em IEEE Trans.
Circuits System-I} \textbf{44(3)}, 268-271.\\
\newpage
\textbf{Figures Captions}:

Fig1: (a) Bifurcation diagram of $\Phi_{2}^{(2)}(x,\alpha)$, for
$\alpha\in(0.5,\infty)$, it is ergodic and for $\alpha\in(0,.5)$,
it has stable fixed point at $x=0$

Fig1: (b) Bifurcation diagram of $\Phi_{3}^{(1)}(x,\alpha)$, where
for $\alpha\in(1/3,3)$, it is ergodic and for $\alpha\in(0,1/3)$,
it has stable fixed point at $x=0 $, while for
$\alpha\in(3,\infty)$, it has stable fixed point $x=1$.

Fig2: (a) Solid surface shows the variation of Lyapunov
characteristic exponent $\Phi_{3,5}^{\alpha_1,\alpha_2}(x)$, in
terms of the parameters $\alpha_1$ and $\alpha_2$.

Fig2: (b) Solid surface shows the variation of Lyapunov
characteristic exponent $\Phi_{4,8}^{\alpha_1,\alpha_2}(x)$, in
terms of the parameters $\alpha^{'}_{1}$ and $\alpha^{'}_{2}$.

Fig3: Block Diagram

Fig4: (a) Plain-image, (b) Ciphered image,(c) and (d) Encryption
with wrong keys.

Fig5: (a) Histogram of plain-image,(b) Histogram of ciphered-image.

Fig6: Correlations of two horizontally adjacent pixels in the
plain-image and in the cipher-image: (a) Correlation analysis of
plain- image,(b) Correlation analysis of cipher-image.

\newpage
\begin{table}
\begin{center}
\caption{Correlation coefficients of two adjacent pixels in two
images} \label{tab:1}
\begin{tabular}{lllllllllllllllll}
\hline\noalign{\smallskip}
 &  &&&&&&Plain image& &&&&&& & & Ciphered image \\
\noalign{\smallskip}\hline\noalign{\smallskip}
Horizontal  &&&&&&& 0.9525 & & &&&&&& & 0.0023\\
vertical    &&&&&&& 0.9443 & &&&&&& & & 0.0026\\
Diagonal    &&&&&&& 0.9066 & &&&&&& & & 0.0013\\
\noalign{\smallskip}\hline
\end{tabular}
\vspace{2cm}
\end{center}
\end{table}
\end{document}